# The interplay between thermodynamics and kinetics in the self-assembly of DNA functionalized nanoparticles

Runfang Mao and Jeetain Mittal*

*Department of Chemical and Biomolecular Engineering,*

*Lehigh University, Bethlehem, PA 18015-4791, USA*

E-mail: jeetain@lehigh.edu

**Abstract**

We use a coarse-grained model of DNA-functionalized particles to understand the role of DNA chain length on their self-assembly. We find that the increasing chain length for a given particle size decreases the propensity to form ordered crystalline assemblies, and instead, disordered structures start to form when the chain length exceeds a certain threshold, which is consistent with the previous experiments. Further analysis of the simulation data suggests weakening interparticle interactions with increasing chain length, thereby shifting the suitable assembly conditions to lower temperatures at which assembly dynamics are unfavorable. This highlights a complex interplay between thermodynamics and dynamics, which we suggest can be modulated by changing the system parameters such as DNA grafting density, resulting in successful crystallization of particles with longer DNA chain lengths. Our results highlight the power of computational modeling in elucidating the fundamental design principles and guiding the assembly of nanoparticles to form complex nanostructures.

The self-assembly of long-range ordered crystal structures from simple building blocks is attracting growing attention owing to its applications in the synthesis of functionalized materials for molecular sensing [1,2], medical diagnosis [3], catalysis [4] and plasmonic materials [5–8]. One particularly powerful approach to assemble complex structures exploits DNA functionalized particles (DFPs) by leveraging carefully tuned interparticle interactions between partially complementary DNA molecules attached to the particle surface [9–13]. However, the formation of long-range ordered crystalline structures remains challenging because DFPs can be easily kinetically trapped into disordered structures [14–17]. Several design rules have become available to guide the selection of appropriate DNA properties such as strand length and grafting density to control the assembly process [11,13,18–25]. Specifically, the DNA strand length is an important design variable to control the interparticle spacing and thereby crystal porosity with potential applications in the field of plasmonics and materials for separation technologies.

However, previous experimental studies suggest a limitation on how much one can vary the strand length in a given system. Macfarlane et al. found that the face-centered-cubic (FCC) crystalline structures in a single-component DFP system with a self-complementary DNA sequence only formed if the DNA strand length is comparable to the nanoparticle diameter [19]. Xiong et al. used a binary DFP system to show that the expected body-centered-cubic (BCC) structure forms over a limited range of linker lengths dependent upon the number of linkers per particle [18]. These results highlight a fundamental restriction on varying the crystal lattice constant by merely changing the DNA strand length [26–30].

The precise nature of this limitation is currently not well understood, which severely limits our fundamental understanding of design rules that underlie assembly of DFPs. The successful assembly of DFPs in crystalline structures relies on a complex interplay balance between both thermodynamic and kinetic factors. The binding free energy between a pair of DFPs, which is quite sensitive to the system parameters and solution conditions such as temperature and salt concentration [31–33], has been identified as an essential indicator of successful crystallization. Using state-of-the-art DFP assembly simulations, Cruz and co-workers demonstrated the role of binding dynamics between a pair of DFPs in determining the crystallinity of self-assembled structures and identified a suitable range of parameters for annealing into ordered lattices [34,35]. Based on these previous studies, we expect that arbitrarily increasing DNA strand length may affect thermodynamics and/or dynamics of DFP assembly in a way that prevents crystallization.

The focus of this paper is to provide a mechanistic understanding of these issues, which should help alleviate this limitation on using specific linker lengths for DFP crystallization. We investigate this using a coarse-grained (CG) model of DFPs that represents each nucleotide with two interaction sites (CG beads) and can successfully capture temperature-dependent melting transitions and other sequence-specific DNA features [36]. First, we identify a suitable range of temperatures for which a system of DFPs assembles into the expected FCC structures in an unbiased molecule dynamics (MD) simulation initialized in a disordered state. Based on extensive simulations, we find that increasing the DNA linker length does indeed prevent the assembly of DFPs into crystalline structures, which is consistent with previous experiments [19]. Importantly, we also find that the underlying reasoning for this behavior is due to the shift in binding free energy as a function of temperature, which necessitates assembly at low temperatures, conditions that are unsuitable from the standpoint of binding dynamics. We propose a strategy to overcome this limitation and demonstrate the successful assembly of DFPs with longer linker lengths into crystalline structures in our MD simulations. Our study, therefore, highlights the critical role that CG models can play in uncovering fundamental rules governing the self-assembly of DFPs, and guiding the realization of complex nanostructures based on DFPs.

To study the role of linker length on the crystallization of DFPs, we use a simple partially self-complementary DNA sequence $T_n$CGCG (Figure 1a), where n is the number of non-hybridizable poly-thymine linker beads. Similar to our previous work [36], each rigid nanoparticle is made up of 100 surface beads that have purely repulsive interactions for each other as well as with the DNA beads. The complementary G:C pairs can interact favorably in our model if the DNA strands involved are conjugated to two different particles. More details about the DFP model can be found in the Methods section and our previous work. We graft each nanoparticle with 32 single-stranded DNA (ssDNA) as shown in Figure 1a and vary the linker length from n = 2 to 32, corresponding to DNA length to particle size ratios ($L_{DNA}/D_p$) of 0.5 to 3.33.

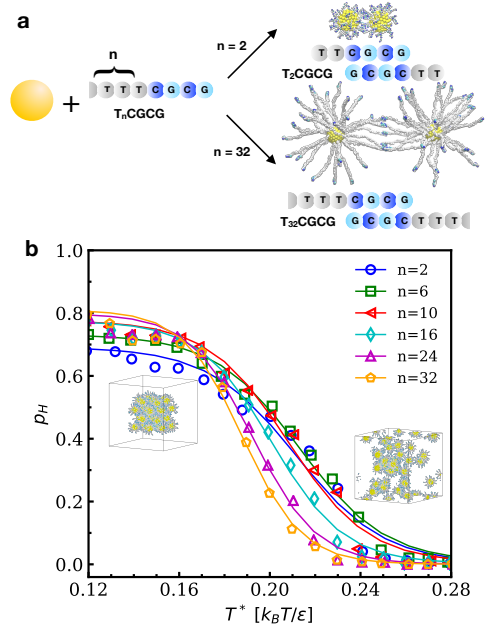

*Figure 1.* (a) Schematic of CG model with DFPs grafted with self-complementary DNA strands, $T_n$CGCG, for different strand lengths used in simulations. Each DNA strand is composed of two regions: the n-base spacers (i.e., number of poly-T beads) and the sticky-end ("CGCG"). The strand lengths are controlled by adjusting the number of spacers from n=2 to 32. Two CG examples with shortest strand $T_2$CGCG and the longest strand $T_{32}$CGCG used in simulations are shown here. (b) Percentage of hybridized DNA complementary pairs $p_H$ as a function of temperature for a different number of spacers, n. The blue, green, red, cyan, purple and orange curves represent n=2,6,10,14,24 and 32 respectively. The insets show the initial and final snapshots of simulations at the lowest ($T^*$=0.12) and the highest temperature ($T^*$=0.28) separately.

The assembly of DFPs is sensitively dependent on temperature [37], with sharp melting transitions observed over a narrow temperature range. To realize ordered assemblies from DFPs in an experiment or simulation, it is critical to identify a suitable temperature or annealing protocol first [38–40]. It was recently shown that suitable assembly conditions allow for dynamic hybridization, which can be quantified using a parameter based on the percentage of hybridized DNA base pairs $p_H$ [34,35]. The expectation is that $p_H$ should be close to 0.5 so that a lot of DNA binding/unbinding events happen, which allows for particles to reorganize in low energy crystalline structures. As described in detail in the methods, we create FCC lattices at dilute conditions (which are the expected structures for this single-component DFP system from the experiments [19]) to estimate $p_H$ as a function of temperature for different linker lengths.

Figure 1b shows the resulting $p_H$ data as a function of reduced temperature $T^*$ for different n values. The highest fraction of hybridized base pairs is observed at the lowest simulation temperature ($T^*$=0.12), and $p_H$ decreases with increasing $T^*$, reaching a zero value at very high temperatures ($T^*$>0.24). The observed melting behavior is used as a guide to select a range of temperature values between $T^*$=0.18 and 0.25, where we expect thermal fluctuations, causing DNA association/dissociation and assembly into ordered structures.

Next, we conduct MD simulations starting from a disordered configuration (generated by randomly placing particles in the simulation box), as shown in Figure 2a (top) using HOOMD-Blue [41,42], which uses graphics processing units (GPUs) to speed-up simulations of large systems like these (total 44,192 particles for n=16 and 32 DFPs). We conduct five independent simulations for each n and $T^*$ to account for the statistical variability. As it is infeasible to employ commonly used order parameters based on structural factors from scattering patterns to assess crystallinity due to the small number of DFPs, we instead rely on the known differences in the pair correlation function (PCF), g(r), between crystalline and disordered functions to see whether crystallization happened during the simulation or not. The g(r) can be used to define a structural similarity index $S(g,g_{ref})$, where $g_{ref}$ is the PCF for the reference crystal structure (FCC in this case) [43].

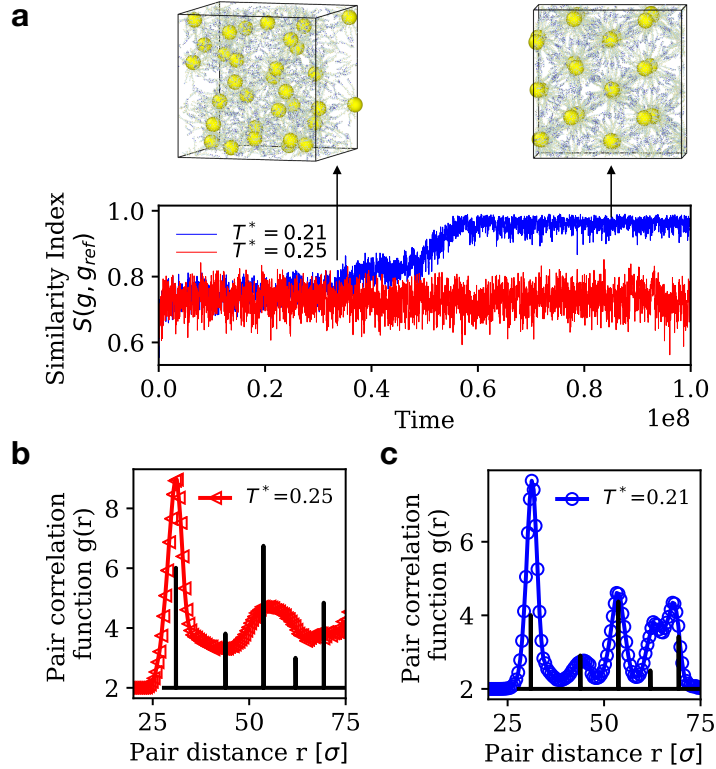

*Figure 2.* (a) Structure similarity index S(g,g$_{ref}$) from comparing reference FCC g(r) and trajectory g(r) along the simulation time at n=10 number of spacers. The red line represents the S(g,g$_{ref}$) obtained at a temperature T$^*$=0.25 where systems form disordered structures. The blue line represents the S(g,g$_{ref}$) at temperature T$^*$=0.21 where the system forms an FCC lattice. The top snapshots show the initial and final configurations at n=10 and T$^*$=0.21. (b) g(r) obtained from final snapshots at n=10 and T$^*$=0.25. (c) g(r) obtained from final snapshots at n=10 and T$^*$=0.21. The black vertical lines in (b) and (c) are the ideal FCC peak locations.

Figure 2a (bottom) shows typical examples of MD trajectories, in terms of S(g,g$_{FCC}$), when crystallization is observed (blue) or not (red) during the simulation. A low value of the similarity index indicates a disordered configuration in this case, whereas a value close to 1 is suggestive of high structural similarity with the reference FCC structure. This disorder and order are clearly depicted through a comparison of the g(r) calculated from the final snapshots for T$^*$=0.25 (Fig. 2b) and T$^*$=0.21 (Fig. 2c), respectively, with the theoretical FCC peak locations. We also looked for the formation of other crystal structures such as hexagonal-close-packed (HCP) and body-centered cubic (BCC) but did not find any evidence of those based on the observed g(r).

Figure 3 summarizes our findings from extensive MD simulations as a function of n and T* in terms of an order diagram. The state points at which successful crystallization is observed during the MD simulation are reported as blue circles; the number inside the circle indicates the number of simulations (out of five), which show the formation of FCC structures. For short DNA strands (i.e., 2<$n$<16), the FCC structures are formed at multiple temperatures, with some cases showing that all five simulations resulted in the formation of ordered structures. For $L_{DNA}/D_p \sim 1$, a significant number of simulations result in crystallization, which is consistent with previous experimental work by Macfarlane et al. [19] As the DNA linker length increases to $n = 24$, however, the crystallization is only observed for one simulation at T*=0.2; further increasing the DNA linker length to $n$=32, only results in the formation of disordered structures. This is consistent with previous experimental studies by Macfarlane et al., which found only disordered structures for $L_{DNA}/D_p > 2$. They proposed that this limitation is due to the slow hybridization and de-hybridization rate of DNA ligands but did not provide guidance on how to overcome this issue.

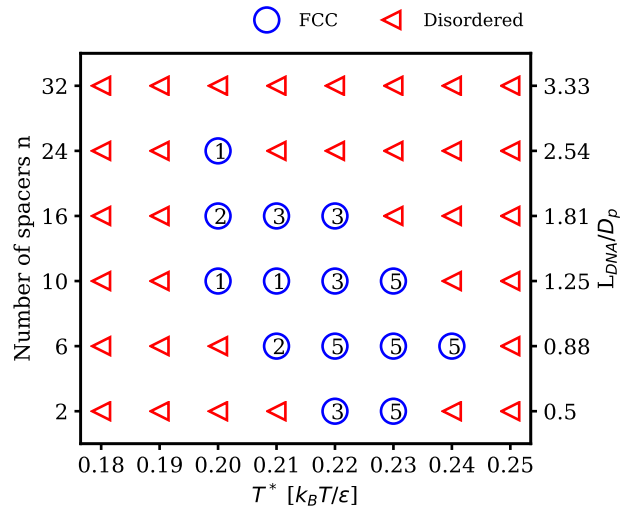

*Figure 3.* Crystallization order diagram as a function of temperature, T*, and the number of spacers, n. Blue circles and red triangles represent the conditions for identifying FCC and disordered structures, respectively. Indices in blue circles show the number of simulations forming FCC lattice from 5 parallel simulations. The second y-axis on the right shows the ratio between DNA strand end-to-end distance and nanoparticle diameter, $L_{DNA}/D_p$. The simulation snapshots and g(r) comparison is shown in SI figure 2 and 3, respectively.

The order diagram (Fig. 3) illustrates two essential features in the linker length-dependent self-assembly: 1) the formation of FCC structures occurs in a narrow temperature range, which is a function of n, and 2) there is a critical linker length (presumably dependent on the particle size [18,19]) above which the particles cannot self-assemble into crystalline structures. To understand this observed behavior, we look for thermodynamic and kinetic factors that have been deemed central to the crystallization of DFPs. Figure 4a shows the binding free energy $\Delta\Delta F$ between a pair of DFPs (see Methods) as a function of temperature for different n values. We highlight with filled symbols the cases for which FCC structures form during any of the simulations in the order diagram. As discussed previously [36], the most effective crystallization conditions likely correspond to the weak-binding regime where the $\Delta\Delta F$ is of the order -1 to -4 $k_BT$ [40], which is essential for the thermal reorganization of the particles. Our simulation data confirms this expectation for most cases except for larger n values for which the formation of FCC structures is not observed, most notably for n=32.

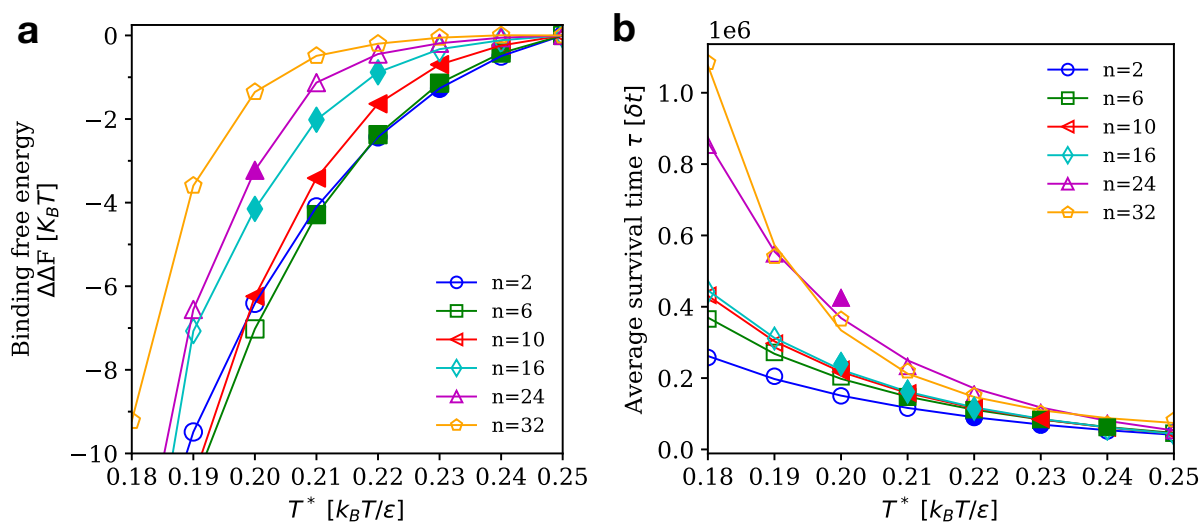

*Figure 4.* (a) DFP binding free energy $\Delta\Delta F$ as a function of temperature for a different number of spacers n. The filled symbols highlight the systems forming an FCC lattice, while the open symbols show the systems forming disordered structures. (b) Average strand survival time

as a function of temperature for a different number of spacers n. The filled symbols highlight the systems forming an FCC lattice, while the open symbols show the systems forming disordered structures.

The results above highlight the need to account for other factors, in addition to the binding free energy, which can explain the observed crystallization behavior for longer linker lengths. Figure 4b shows the average survival time τ, which quantifies the time it takes for DNA strands to dissociate from a hybridized state (see Methods). The state points for which we observe the formation of FCC structures (filled symbols) show faster dynamics with shorter τ values. We notice a rapid increase in τ for longer strands( n=24 and 32) at low $T^*$ values. Again, τ by itself is insufficient to explain the observed crystallization behavior as we do not necessarily see the formation of FCC structures for low τ values except in the cases of the shortest spacers.

We hypothesize that the observed trends in Fig. 3 can be explained by a combination of the two factors (ΔΔF and τ) used in Fig. 4, as we discuss next. The ΔΔF curves shift to the left as a function of $T^*$ with increasing n, which suggests that the binding between a pair of DFPs weakens with increasing linker length (see SI Figure S4) even though the complementary part of the DNA (and the hybridization free energy between a pair of DNA molecules) is unchanged. At first glance, this may seem a bit surprising, but it makes sense if one considers the density of complementary (sticky) parts of the DNA strands around the particle with increasing n. Even though the actual number of DNA strands grafted on the particles remains the same, the effective density of the sticky part decreases with increasing the linker length, as these ends are farther away from the particle surface when hybridization happens between two DFPs. This is also consistent with the melting behavior shown in Fig. 1.

Therefore, one would expect that for longer linkers, the favorable thermodynamic conditions for DFP self-assembly in terms of ΔΔF (weak-binding regime) should be shifted to lower temperatures, conditions that are not suitable from a kinetic standpoint (high τ values). This interesting interplay between thermodynamic and kinetic factors is, therefore, likely responsible for the observed behavior here and in previous experiments. This finding is of significant importance for guiding the assembly of DFPs and other types of nanoparticles/colloids, as it is often tempting/simpler to focus on either thermodynamic or kinetic factors in isolation. However, the results here suggest that the actual assembly may depend on both of these factors in a more complicated way, which is difficult to discern a priori.

To validate our findings that weak-binding, coupled with faster dynamics, will lead to successful crystallization of DFPs, we propose to modify the system parameters that can accomplish this for longer linker lengths for which we were unable to observe the formation of FCC structures during the simulation times. As discussed earlier, the effective density of sticky DNA ends goes down with increasing n, causing an increase in $\Delta\Delta F$. If we can manage to reverse this trend in $\Delta\Delta F$ for a fixed linker length, so that the favorable thermodynamic conditions for assembly overlap with favorable dynamics in terms of temperature, it should result in the formation of FCC structures for longer linker lengths. The most straightforward approach to accomplish this computationally, and in all likelihood experimentally with some additional effort on synthesis, is to increase the number of DNA strands grafted on the particle surface.

We conduct additional simulations to test our hypothesis directly and present the results in Figure 5 as a function of the number of DNA strands attached to a particle $N_s$ and temperature for $n=24$. We find the formation of FCC structures in many more simulations for higher $N_s$ values, thereby confirming the synergistic coupling between $\Delta\Delta F$ and $\tau$ (Figs. 5c,d). As expected, the $\Delta\Delta F$ curves shift toward higher temperatures with increasing $N_s$ for fixed n. This, in turn, shifts the favorable assembly conditions (weak-binding regime) to higher temperatures at which the DNA hybridization dynamics are also more favorable, though the $\tau$ itself is mostly independent of $N_s$ in this case.

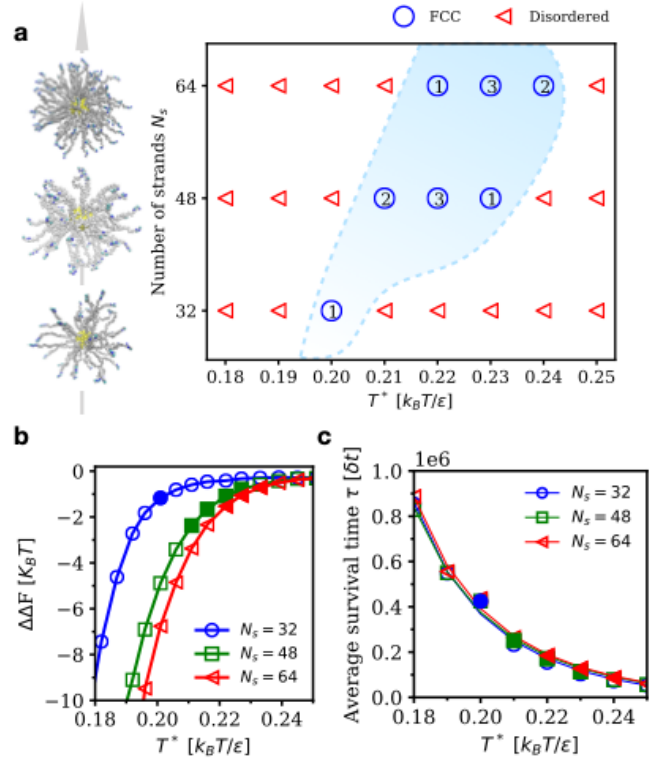

*Figure 5.* (a) Schematic representation for increasing number of strands $N_s$ around nanoparticles and the resulting crystallization order diagram as a function of temperature and number of strands $N_s$ at n=24. Blue circles represent the conditions forming an FCC lattice while red triangles represent systems forming disordered structures. The indices in blue circles show the total number of simulations resulting in an FCC lattice among 5 parallel simulations. The filled area represents the conditions favoring crystallization. (b) DFP binding free energies as a function of temperature for different numbers of strands $N_s$ at n=24. The filled symbols highlight the systems forming FCC lattices, while the open symbols show the systems forming disordered structures. (c) Average strand survival time as a function of temperature for different numbers of strands $N_s$ at n=24. The filled symbols highlight the systems forming FCC lattices, while the open symbols show the systems forming disordered structures.

In conclusion, we use a coarse-grained model of DNA-functionalized particles to identify if there is a fundamental limitation in assembling such particles in crystalline structures with increasing DNA chain length. Indeed, we do observe in our molecular dynamics simulations that the propensity to form ordered structures decreases with increasing chain length. Our analysis suggests that these results can be explained in terms of weakened interparticle interactions or a shift in the melting transition to lower temperatures with increasing chain length, which forces assembly at temperatures for which DNA hybridization kinetics are not favorable. We identified a simple design strategy based on the number of DNA strands grafted on a particle that can be leveraged to facilitate assembly into crystalline structures. Our results should be extensible to the experiment as the computational model faithfully represents the fundamental details of these systems. The insights from this study concerning the coupled role of thermodynamics and kinetics, which to our knowledge has not been reported previously, may be broadly applicable to the self-assembly of other systems as well.

## Methods and simulation details

**Coarse-grained model**

The DFPs are modeled as core nanoparticles with grafted single-stranded DNA molecules (ssDNA) using the model from Ding et al. [36]. The nanoparticle is represented as a rigid hollow sphere of radius 5σ that is consists of 100 surface beads, and the surface beads are uniformly distributed around a central ghost bead. The ssDNA sequence is selected to be self-complementary ($T_n$CGCG), where n represents the number of spacer poly-Thymine beads. The DNA strand length is varied by changing the number of spacers while the sticky part is kept the same as CGCG in all the simulations. Each particle is functionalized with 32 ssDNA molecules (unless mentioned otherwise) that are uniformly attached to the surface beads. The base pairs are covalently linked and attached to a nanoparticle surface using finitely extensible nonlinear elastic (FENE) potential with separation a distance $R_0$ = 1.5 σ.

**Estimating the percentage of hybridized complementary DNA pairs $p_H$**

First, the periodic FCC lattices are created with 32 explicitly modeled DFPs frozen in space with their separation distances adjusted for different strand lengths (see SI Fig S1). Then, the DNA strands are fully relaxed at the lowest simulation temperature ($T^*$=0.12) for a total of 1e8 simulation time steps. After DNA strands are fully relaxed, the central core particles are allowed to move freely, and the $p_H$ is estimated from the heating simulations, which involve gradually increasing the temperature from T*=0.12 to T*=0.28 for a total of 1e8 time steps.

**Molecular dynamics simulation details**

We conduct molecular dynamic (MD) simulations to study the multi-particle assembly starting from a disordered DFP configuration. To enhance simulation efficiency, we use the HOOMD-blue package to perform MD simulations on Graphics processing units (GPUs). All the simulations are performed in a canonical ensemble (fixed NVT) using the Langevin thermostat. Periodic boundary conditions are applied to all three dimensions to minimize the boundary effects. The assembly simulations are started from an initially disordered configuration generated by placing DFPs randomly in the simulation box. The total simulation time is 3x10$^8$ with time step $\delta t = 0.05 \left(\frac{\varepsilon}{m\sigma^2}\right)^{-\frac{1}{2}}$. In total, 32 explicitly modeled DFPs are added into the simulation box at packing fraction around η = 0.65. Packing fraction is defined from η = (4π/3)(R/L)³, where R is the effective DFP radius (half of D$_s$) approximated from the separation distance as shown in SI Fig S3 and L is the simulation box length used in each direction. All quantities are presented in Lennard-Johns (LJ) reduced units, and all particle beads have equal mass, m = 1. The reduced temperature T* is measured in terms of $k_B T/\varepsilon$.

**Structure similarity index**

Given two structures, α and β and their radial distribution functions g$_α$(r) and g$_β$(r), one can define a similarity parameter as follows:

$$S(g_\alpha, g_\beta) = \frac{\int_0^{r_c} g_\alpha(r) g_\beta(r)\, dr}{\sqrt{\int_0^{r_c} g_\alpha^2(r)\, dr} \sqrt{\int_0^{r_c} g_\beta^2(r)\, dr}}$$

Here, S is expected to vary from 0 to 1, where 0 corresponds to two completely dissimilar structures and 1 in the case of two identical structures. The similarity index S(g,g$_{ref}$) was calculated from the simulation trajectory by using the reference g (r) from the equilibrium FCC lattice.

**Binding free energy calculations**

Replica exchange molecular dynamics (REMD) simulations are performed using LAMMPS to calculate the potential of mean force (PMF) between two DFPs similar to our previous work. [36] These simulations are conducted in a canonical ensemble, in which the temperature is maintained using a Langevin thermostat with damping parameter $\tau = 1\left(\frac{\varepsilon}{m\sigma^2}\right)^{-\frac{1}{2}}$. The simulation box length is set to at least three times the effective DFP diameter (nanoparticle core diameter plus twice the ssDNA contoured length) in all 3-dimensions with periodic boundary conditions. In REMD simulations, 32 temperature replicas are used to allow for sufficient exchange probability. The total simulation time is 1e9 steps with a step size $\delta t = 0.01\left(\frac{\varepsilon}{m\sigma^2}\right)^{-\frac{1}{2}}$. The first 1e8 time steps per replica are discarded for equilibration.

The potential of mean force (PMF) as a function of pair distance is calculated as $\Delta F(r) = -k_B T \ln[g(r)] + c$, where $g(r)$ is the radial distribution function, $k_B$ is the Boltzmann constant and $c$ is an additive constant. The DFP binding free energy between bound state and unbound state $\Delta\Delta F$ is calculated as $\Delta\Delta F = \Delta F_{bound} - \Delta F_{unbound} = -k_B T \ln(\int_0^{r_{cut}} e^{-\beta F(r)} dr / \int_{r_{cut}}^{\infty} e^{-\beta F(r)} dr)$, where $r_{cut}$ is the distance up to which the particles are considered bound and $r_{max}$ is the distance at which the $g(r)$ has plateaued to its bulk value.

**Survival time calculations**

To estimate DNA strand hybridization dynamics, we record all bound DNA strands at time t=0 and track the hybridization state throughout the simulation trajectory. A pair of DNA strands are considered bound if more than 50% of the sticky base pairs are hybridized, otherwise these strands are considered unbound. The hybridization fraction f(H) follows an exponential decay, as shown in SI Figure S5. The average strand survival time $\tau$ can be estimated from equation f(H) = Aexp(−t/$\tau$)+f(H)$_0$.

# Acknowledgment

This work is supported by the US Department of Energy, Office of Science, Basic Energy Sciences Award DE-SC00013979. This research used resources of the National Energy Research Scientific Computing Center, a DOE Office of Science User Facility supported under Contract No. DE-AC02-05CH11231. Use of the high-performance computing capabilities of the Extreme Science and Engineering Discovery Environment (XSEDE), which is supported by the National Science Foundation, project no. TG-MCB120014 is also gratefully acknowledged.

# Reference

[1]   Y. D. Park, A. T. Hanbicki, S. C. Erwin, C. S. Hellberg, J. M. Sullivan, J. E. Mattson, T. F. Ambrose, A. Wilson, G. Spanos, and B. T. Jonker, Science (80-. ). (2002).
[2]   L. Lin, Y. Liu, L. Tang, and J. Li, Analyst (2011).
[3]   C. A. Mirkin and G. B. Rathmann, in *MRS Bull.* (2010).
[4]   Y. Kang, X. Ye, J. Chen, Y. Cai, R. E. Diaz, R. R. Adzic, E. A. Stach, and C. B. Murray, J. Am. Chem. Soc. (2013).
[5]   D. S. Sebba, J. J. Mock, D. R. Smith, T. H. LaBean, and A. A. Lazarides, Nano Lett. (2008).
[6]   D. J. Park, C. Zhang, J. C. Ku, Y. Zhou, G. C. Schatz, and C. A. Mirkin, Proc. Natl. Acad. Sci. U. S. A. (2015).
[7]   K. L. Young, M. B. Ross, M. G. Blaber, M. Rycenga, M. R. Jones, C. Zhang, A. J. Senesi, B. Lee, G. C. Schatz, and C. A. Mirkin, Adv. Mater. (2014).
[8]   A. Kuzyk, R. Schreiber, Z. Fan, G. Pardatscher, E. M. Roller, A. Högele, F. C. Simmel, A. O. Govorov, and T. Liedl, Nature (2012).
[9]   E. W. Gehrels, W. B. Rogers, and V. N. Manoharan, Soft Matter (2018).
[10]  C. A. Mirkin, R. L. Letsinger, R. C. Mucic, and J. J. Storhoff, Nature (1996).
[11]  D. Nykypanchuk, M. M. Maye, D. Van Der Lelie, and O. Gang, (n.d.).
[12]  A. P. Alivisatos, K. P. Johnsson, X. Peng, T. E. Wilson, C. J. Loweth, M. P. Bruchez, and P. G. Schultz, Nature (1996).
[13]  R. J. Macfarlane, B. Lee, M. R. Jones, N. Harris, G. C. Schatz, and C. A. Mirkin, Science (80-. ). (2011).
[14]  L. Di Michele, F. Varrato, J. Kotar, S. H. Nathan, G. Foffi, and E. Eiser, Nat. Commun. **4**, (2013).
[15]  A. J. Kim, R. Scarlett, P. L. Biancaniello, T. Sinno, and J. C. Crocker, Nat. Mater. **8**, 52 (2009).
[16]  Y. Wang, Y. Wang, X. Zheng, É. Ducrot, M. G. Lee, G. R. Yi, M. Weck, and D. J. Pine, J. Am. Chem. Soc. **137**, 10760 (2015).
[17]  C. Knorowski, A. Travesset, R. J. Macfarlane, C. A. Mirkin, M. O. de la Cruz, C. A. Mirkin, C. A. Mirkin, H. Jinnai, Y. Wu, D. Poulsen, J. M. J. Frechet, A. P. Alivisatos, and T. Xu, Soft Matter **8**, 12053 (2012).

[18] H. Xiong, D. Van Der Lelie, and O. Gang, Phys. Rev. Lett. (2009).
[19] R. J. Macfarlane, M. R. Jones, A. J. Senesi, K. L. Young, B. Lee, J. Wu, and C. A. Mirkin, Angew. Chemie - Int. Ed. **49**, 4589 (2010).
[20] R. J. Macfarlane, R. V. Thaner, K. A. Brown, J. Zhang, B. Lee, S. T. Nguyen, and C. A. Mirkin, Proc. Natl. Acad. Sci. (2014).
[21] O. Gang and A. V Tkachenko, 381 (2018).
[22] F. Vargas Lara and F. W. Starr, Soft Matter (2011).
[23] O. Padovan-Merhar, F. V. Lara, and F. W. Starr, J. Chem. Phys. (2011).
[24] S. Angioletti-Uberti, B. M. Mognetti, and D. Frenkel, Phys. Chem. Chem. Phys. **18**, 6373 (2016).
[25] A. Seifpour, S. R. Dahl, B. Lin, and A. Jayaraman, Mol. Simul. (2013).
[26] H. D. Hill, R. J. Macfarlane, A. J. Senesi, B. Lee, S. Y. Park, and C. A. Mirkin, Nano Lett. (2008).
[27] J. J. Storhoff, R. Elghanian, R. C. Mucic, C. A. Mirkin, and R. L. Letsinger, J. Am. Chem. Soc. **120**, 1959 (1998).
[28] C. A. Elghanian, Robert;Storhoff, James J;Mucic, Robert C;Letsinger, Robert L;Mirkin, Science (80-. ). **277**, 1078 (1997).
[29] M. Paule Pileni, New J. Chem. (1998).
[30] J. R. Heath, C. M. Knobler, and D. V. Leff, J. Phys. Chem. B (1997).
[31] R. J. Macfarlane, M. R. Jones, B. Lee, E. Auyeung, and C. A. Mirkin, Science (80-. ). (2013).
[32] S. Y. Park, A. K. R. Lytton-Jean, B. Lee, S. Weigand, G. C. Schatz, and C. A. Mirkin, Nature **451**, (2008).
[33] M. R. Jones, N. C. Seeman, and C. A. Mirkin, Nanomaterials **347**, (2015).
[34] T. I. N. G. Li, R. Sknepnek, R. J. Macfarlane, C. A. Mirkin, and M. Olvera De La Cruz, Nano Lett. (2012).
[35] T. I. N. G. N. G. Li, R. Sknepnek, and M. O. De La Cruz, J. Am. Chem. Soc. (2013).
[36] Y. Ding and J. Mittal, J. Chem. Phys. **141**, 184901 (2014).
[37] M. Song, Y. Ding, M. A. Snyder, and J. Mittal, Langmuir **32**, 10017 (2016).
[38] M. Song, Y. Ding, H. Zerze, M. A. Snyder, and J. Mittal, Langmuir (2018).
[39] M. T. Casey, R. T. Scarlett, W. Benjamin Rogers, I. Jenkins, T. Sinno, and J. C. Crocker, Nat. Commun. (2012).
[40] R. Dreyfus, M. E. Leunissen, R. Sha, A. Tkachenko, N. C. Seeman, D. J. Pine, and P. M. Chaikin, Phys. Rev. E - Stat. Nonlinear, Soft Matter Phys. (2010).
[41] J. A. Anderson, C. D. Lorenz, and A. Travesset, J. Comput. Phys. (2008).
[42] J. Glaser, T. D. Nguyen, J. A. Anderson, P. Lui, F. Spiga, J. A. Millan, D. C. Morse, and S. C. Glotzer, Comput. Phys. Commun. (2015).
[43] E. Pretti, H. Zerze, M. Song, Y. Ding, R. Mao, and J. Mittal, Sci. Adv. **5**, eaaw5912 (2019).

## *Supporting information (SI) figures*

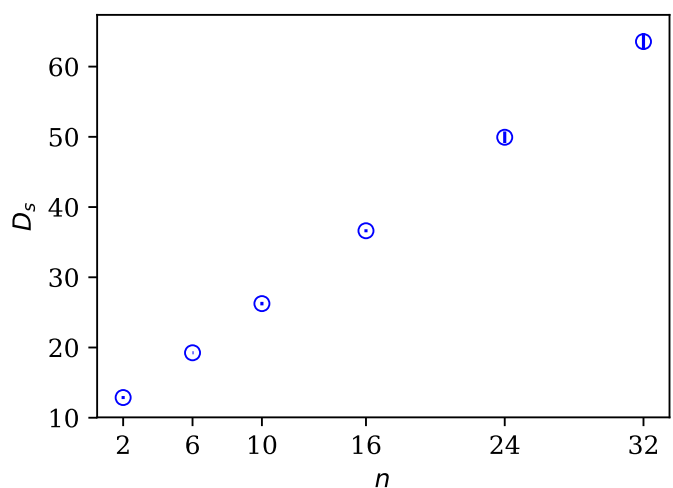

Fig S1. The separation distance, $D_s$, as a function of the number of spacers estimated from the first peak of the pair correlation function, $g(r)$, from molecular dynamics self-assembly simulations.

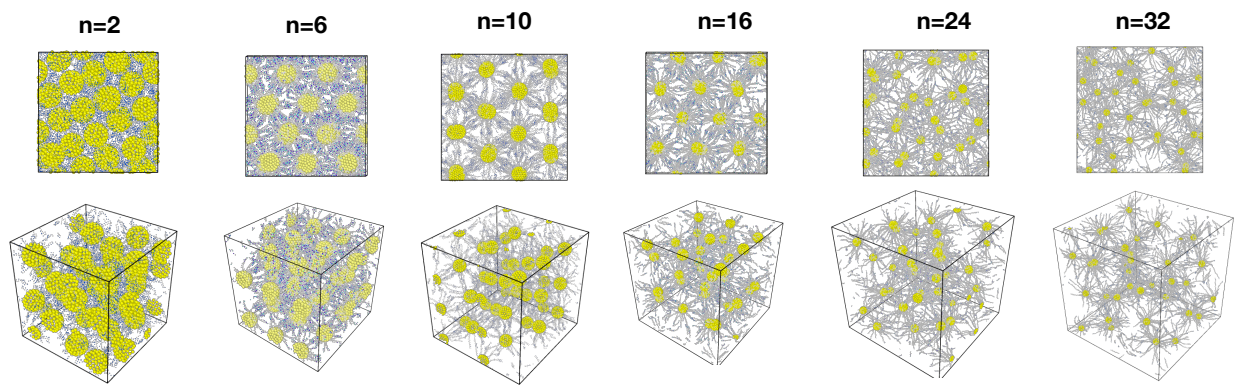

Fig S2. Simulation snapshots for different strand lengths with the number of spacers from n=2 to 32 at $T^*=0.22$.

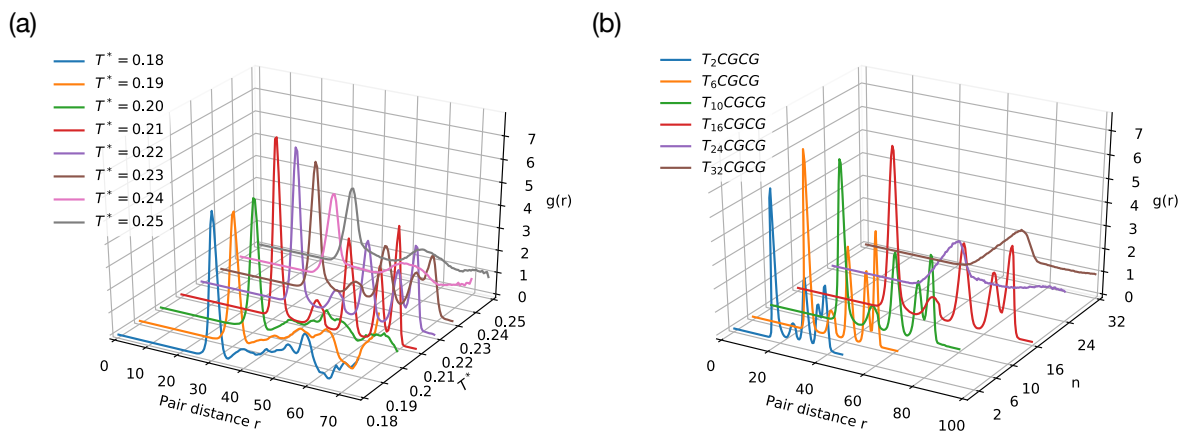

Fig S3. (a) Pair correlation function, g(r), obtained from the average over final few snapshots for the spacer length n=10 at different temperatures. (b) g(r) obtained from the average over final few snapshots for different spacer lengths, n, at $T^*$=0.22.

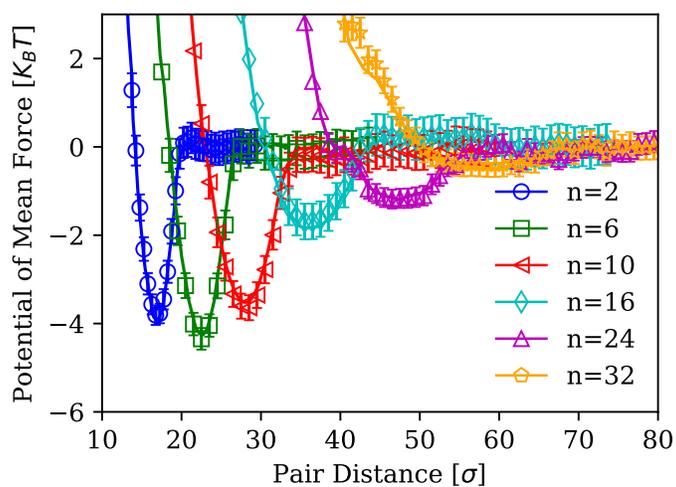

Fig S4. Potential of mean force (PMFs) for different strand lengths obtained from replica exchange molecular dynamics (REMD) simulations.

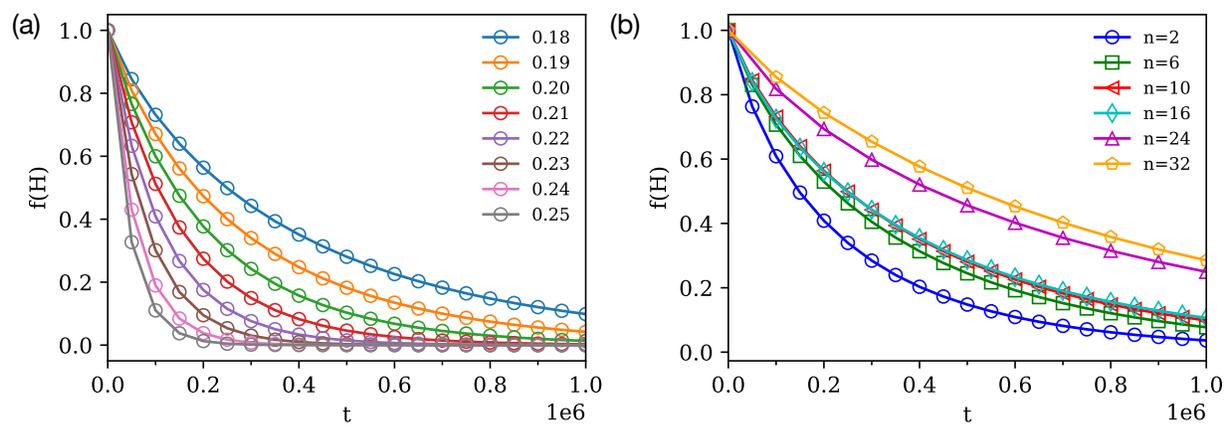

Fig S5. (a) The fraction of hybridization, f(H), as a function of simulation time t for the spacer length n=10 at different temperatures (see legend). (b) f(H) as a function of simulation time for different spacer lengths at $T^*$=0.18.